\documentclass[conference]{IEEEtran}
\IEEEoverridecommandlockouts
\usepackage{cite}
\usepackage{amsmath,amssymb,amsfonts}
\usepackage{algorithmic}
\usepackage{graphicx}
\usepackage{textcomp}
\usepackage{xcolor}
\usepackage{multirow}

\def\BibTeX{{\rm B\kern-.05em{\sc i\kern-.025em b}\kern-.08em
    T\kern-.1667em\lower.7ex\hbox{E}\kern-.125emX}}
\begin{document}

\title{uTalk: Bridging the Gap Between Humans and AI\\
}

\author{
\IEEEauthorblockN{
Hussam Azzuni\IEEEauthorrefmark{1}\textsection,
Sharim Jamal\IEEEauthorrefmark{1}\textsection,
and Abdulmotaleb Elsaddik\IEEEauthorrefmark{1}\IEEEauthorrefmark{2}}

\IEEEauthorblockA{\IEEEauthorrefmark{1}Mohamed Bin Zayed University of Artificial Intelligence\\
Masdar City, Abu Dhabi, UAE\\
Email: \{hussam.azzuni, sharim.jamal\}@mbzuai.ac.ae}

\IEEEauthorblockA{\IEEEauthorrefmark{2}School of Electrical Engineering and Computer Science\\
University of Ottawa\\
Ottawa, ON, Canada K1N 6N5\\
Email: elsaddik@uottawa.ca}
}

\maketitle
\begingroup\renewcommand\thefootnote{\textsection}
\footnotetext{Equal contribution}
\endgroup

\begin{abstract}
Large Language Models (LLMs) have revolutionized various industries by harnessing their power to improve productivity and facilitate learning across different fields. One intriguing application involves combining LLMs with visual models to create a novel approach to Human-Computer Interaction. The core idea of this system is to create a user-friendly platform that enables people to utilize ChatGPT's features in their everyday lives. uTalk is comprised of technologies like Whisper, ChatGPT, Microsoft Speech Services, and the state-of-the-art (SOTA) talking head system SadTalker. Users can engage in human-like conversation with a digital twin and receive answers to any questions. Also, uTalk could generate content by submitting an image and input (text or audio). This system is hosted on Streamlit, where users will be prompted to provide an image to serve as their AI assistant. Then, as the input (text or audio) is provided, a set of operations will produce a video of the avatar with the precise response. This paper outlines how SadTalker's run-time has been optimized by 27.69\% based on 25 frames per second (FPS) generated videos and 38.38\% compared to our 20FPS generated videos. Furthermore, the integration and parallelization of SadTalker and Streamlit have resulted in a 9.8\% improvement compared to the initial performance of the system. 

\end{abstract}

\begin{IEEEkeywords}
Interactive System, Conversational AI, Content Creation, Human-Computer Interaction, Digital Twins, LLM, User Experience
\end{IEEEkeywords}

\section{Introduction}

Large language models (LLMs) such as OpenAI's ChatGPT \cite{chatgpt}, Google's Bard, and Meta's LLaMa \cite{touvron2023llama} have revolutionized the way we access information. These models leverage extensive data training to perform various tasks such as recognition, generation, translation, and summarization. Different approaches for disseminating this knowledge are necessary. For example, SpeechGPT \cite{zhang2023speechgpt} is an LLM with inherent cross-modal conversational skills. AudioGPT \cite{huang2023audiogpt} enhances LLMs to process complex audio information and engage in spoken conversations. FaceChat \cite{alnuhait2023facechat} is a promising approach to creating dynamic face-to-face conversation utilizing ChatGPT. Our approach is similar as we both tackle avatar-based ChatGPT for a more engaging conversation. However, our method focuses on optimizing and integrating SadTalker with various software, towards achieving a user-friendly implementation streamlining the information retrieval process. This paper presents a comprehensive system that combines various Application Programming Interface (API) software and open-source codes to create an interactive virtual avatar capable of engaging in informative conversations. The system follows a multi-step process: First, Whisper API allows the transcription of the user's spoken question, while having the option to provide a text input based on their preference. Then, the obtained text and a predetermined prompt are fed into ChatGPT. The additional prompt ensures the precision of the answers. The response generated by ChatGPT is converted into speech using the Speech Studio within Azure Cognitive Services. Finally, the resulting audio and the selected avatar image are fed to SadTalker to create a digital twin \cite{el2018digital} in the form of a talking portrait answering the user's question. Alternatively, the input, whether text or audio, could be given directly to generate videos for content creation. Our framework provides an immersive experience that enables users to engage in personalized conversations with avatars of their choice. Additionally, it can be utilized for creating chatbots, automating customer service interactions, building educational platforms, developing personal assistants, and creating translation tools that break down language barriers. Our contributions can be summarized as follows:

\begin{itemize}
    \item Framework with an integrated and optimized SadTalker and APIs for Human-Computer interactive avatar.
    \item Optimizing the run-time of SadTalker by \(\approx 27.69\%\)
    \item User Interface (UI) offering audio and text inputs for conversing with AI or creating content.
    \item Providing FPS adjustment feature for video generation.
    \item Applying the context of the two previous questions and answers enhances the user experience. 
\end{itemize}

\begin{figure*}[t]
\centerline{\includegraphics[width=0.97\textwidth]{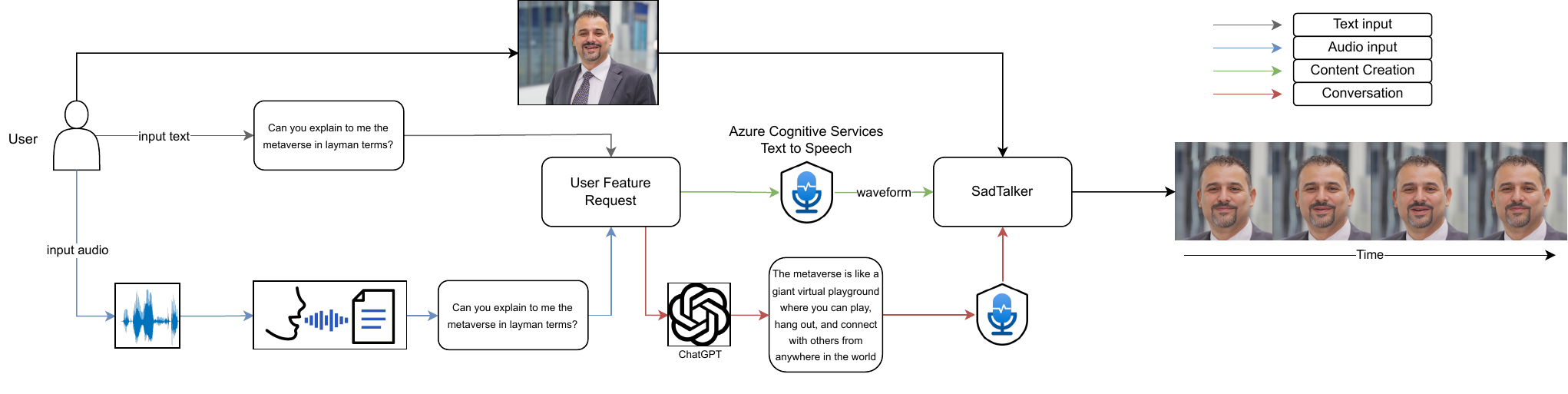}}
\caption{uTalk pipeline consists of two main features: Content Creation and conversing with a digital twin. Both processes begin with the user providing an input, either in the form of audio or text. The input is then processed generating a waveform that is given to SadTalker to generate the video.}
\label{uTalk}
\end{figure*}

\section{Profiling the overall script}

Profiling the script is essential to identify code bottlenecks and improve its performance and speed. In this case, cProfile identifies the main time-consuming blocks within the code. The elapsed times may overlap and not add up to the overall time. The time needed to generate a video given a 7-second audio clip is 40.59 seconds, based on a NVIDIA Geforce RTX 3090Ti and a 12th Gen i9-12900 Intel CPU. The list below shows the bottlenecks with the number of calls written afterward:

\begin{itemize}
    \small
    \item animate.py (generate) = 34.95s (1)
    \item mimwrite = 24.30s (2)
    \item face-enhancer.py = 23.94s (176)
    \item utils.py (enhance) = 22.80s (175)
    \item face-enhancer.py (paste-faces-to-input-images) = 12.53s (175)
    \item make-animation.py = 9.86s (1)
    \item Gaussian blur = 3.75s (350)
\end{itemize}

During cProfile's analysis, multiple bottlenecks are identified within the system. First, optimizing some functions is crucial as it is performed for every frame, creating a substantial overall bottleneck. This is achieved by removing unnecessary lines of code that do not contribute majorly to the final result. Second, "mimwrite" is time-consuming, so we replaced it with a more efficient function from OpenCV.

\section{Framework}

uTalk is a Human-Computer Interaction (HCI) system that utilizes algorithms to create videos based on an image and a statement provided by the user. Fig. 1 shows an abstract view of the complete framework. Initially, the user is prompted to provide an image that will serve as their AI avatar. Subsequently, the user has two options: Users can ask a question via audio or text to receive a response from ChatGPT. Alternatively, the audio or text input provided can be used directly to generate an avatar-speaking video. The whole process is hosted on Streamlit, allowing users to watch the generated videos easily. As the system is comprised of multiple software components, we will carefully explain each component and any necessary software adjustments.

\subsection{Speech to Text}

Whisper API  \cite{radford2022robust}, developed by OpenAI, is an Automated Speech Recognition (ASR) system that transcribes audio files into text. It comes in various sizes, ranging from whisper-tiny, with 39 million parameters, to whisper-large-v2, with 1550 million parameters. Our system uses whisper-tiny to expedite the transcription process. We have incorporated a minimum requirement of two words to be obtained from whisper API to avoid silent inputs. If such an issue occurs, an error message is displayed to alert the user. 

\subsection{ChatGPT and Prompt Engineering}

Our system uses ChatGPT API, specifically GPT 3.5 text-davinci-003 model, for answering questions. We provide an additional prompt to ensure brief responses, which helps the LLM avoid over-explaining, improving the overall speed. Additionally, we maintained context throughout the conversation, ensuring a coherent and meaningful exchange of information.

\subsection{Text to Speech}

uTalk utilizes a text-to-speech (TTS) library within the Speech Studio component in Azure Cognitive Services. This library supports 140 languages and dialects, making it convenient for both English and non-English speakers.

\begin{table*}
\caption{This is a comparison of run-time between SadTalker and our modifications. Results are based on a 5-Run average and a 7-second audio clip. The results provided are a percentage representing the reduction in run-time.}

\resizebox{2\columnwidth}{!}{

\begin{tabular}{|l|cccc|c|l|}

\hline
\multicolumn{1}{|c|}{\multirow{2}{*}{\textbf{Model}}} & \multicolumn{4}{c|}{\textbf{Modifications}} & \multirow{2}{*}{\textbf{FPS}} & \multirow{2}{*}{\textbf{Run-time (seconds)}} \\ \cline{2-5}
\multicolumn{1}{|c|}{} & \multicolumn{1}{c|}{Tqdm Removal} & \multicolumn{1}{c|}{Facexlib Optimization} & \multicolumn{1}{c|}{Removed Intermediate Values} & Replace Mimsave &  &  \\ \hline

SadTalker & \multicolumn{1}{c|}{X} & \multicolumn{1}{c|}{X} & \multicolumn{1}{c|}{X} & X & 25 & 40.637 $\pm$ 0.320  \\ \hline

Proposed mod1 & \multicolumn{1}{c|}{\checkmark} & \multicolumn{1}{c|}{X} & \multicolumn{1}{c|}{X} & X & 25 & 39.930 $\pm$ 0.116 (\textcolor{red}{\textbf{-1.74\%}})\\ \hline

Proposed mod2 & \multicolumn{1}{c|}{\checkmark} & \multicolumn{1}{c|}{\checkmark} & \multicolumn{1}{c|}{X} & X & 25 & 31.182 $\pm$ 0.526 (\textcolor{red}{\textbf{-23.27\%}})  \\ \hline

Proposed mod3 & \multicolumn{1}{c|}{\checkmark} & \multicolumn{1}{c|}{\checkmark} & \multicolumn{1}{c|}{\checkmark} & X & 25 & 31.438 $\pm$ 0.579 (\textcolor{red}{\textbf{-22.64\%}}) \\ \hline

Proposed mod4 (\textcolor{red}{uTalk}) & \multicolumn{1}{c|}{\checkmark} & \multicolumn{1}{c|}{\checkmark} & \multicolumn{1}{c|}{\checkmark} & \checkmark & 25 & 29.385 $\pm$ 0.284 (\textcolor{red}{\textbf{-27.69\%}}) \\ \hline

\end{tabular}}
\label{speedcomparison}
\end{table*}

\subsection{SadTalker}
SadTalker \cite{zhang2023sadtalker} is a state-of-the-art system that produces realistic talking head videos by combining image and audio inputs. It comprises multiple components, including ExpNet, which can precisely capture facial expressions from audio signals, and PoseVAE, which can generate head movements. The final video is created by mapping the 3D motion coefficients to 3D facial key points. SadTalker is the major bottleneck within uTalk; thus, we modified its code and improved its integration within the system to improve its run-time.

\subsubsection{\textbf{Removing Redundant Code}}

The first step towards optimizing our model is the removal of unnecessary libraries or code that does not contribute to the overall performance. We began by removing the tqdm library and its integration in the code base, as it only served to monitor progress. Furthermore, we removed all intermediate values saved and loaded during the four stages. 

\subsubsection{\textbf{Adjustable FPS}}

The code initially generates videos with 25FPS. However, we added adjustable FPS output as a feature to reduce the output FPS without reducing the user experience. A subjective study shows the opinion of 29 participants about nine generated videos with varying FPS from 16 to 24. The study's primary objective is to measure how people perceive the smoothness of AI-generated videos with different FPS.

\subsubsection{\textbf{Enhancing Facexlib Efficiency}}

SadTalker utilizes various models with various input sizes, resizing between 256x256 and 512x512 as deemed necessary, introducing an overhead as resizing hundreds of frames is time-consuming. This is solved by modifying the code to ensure that "FaceRestoreHelper" resizes the code to 256x256 instead of the original 512x512, significantly reducing the run-time without compromising the video quality. We also eliminated functions that did not noticeably impact the final performance, such as adding Gaussian noise for each frame, causing overhead run-time.

\subsubsection{\textbf{Integration of SadTalker with Streamlit}}

The integration between SadTalker and Streamlit is essential for optimizing the system's performance. We have successfully separated the operations into two isolated modules, one for initializing the models and the second for video generation. The modular code is then seamlessly integrated into our Streamlit app. We then take advantage of Streamlit's caching capabilities to pre-load the initialization of models along with the Whisper and ChatGPT API right when the Streamlit web app is loaded for the first time. This approach ensures that these heavy tasks are executed only once and cached for future use. As a result, we have cut down on both response time and computational load, making the system faster and more resource-efficient.

\subsection{User Interface}

\begin{figure*}
    \centering
    \begin{minipage}[t]{0.47\textwidth}
        \fbox{\includegraphics[width=0.9\textwidth]{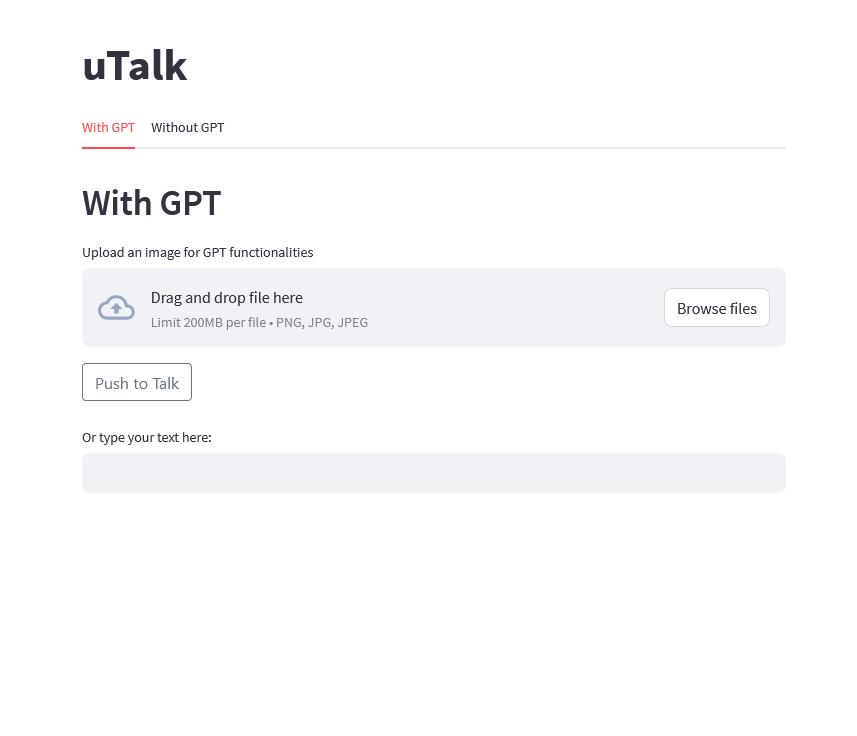}}
    \end{minipage}
    \hspace{0.03\textwidth} 
    \begin{minipage}[t]{0.47\textwidth}
        \fbox{\includegraphics[width=0.9\textwidth]{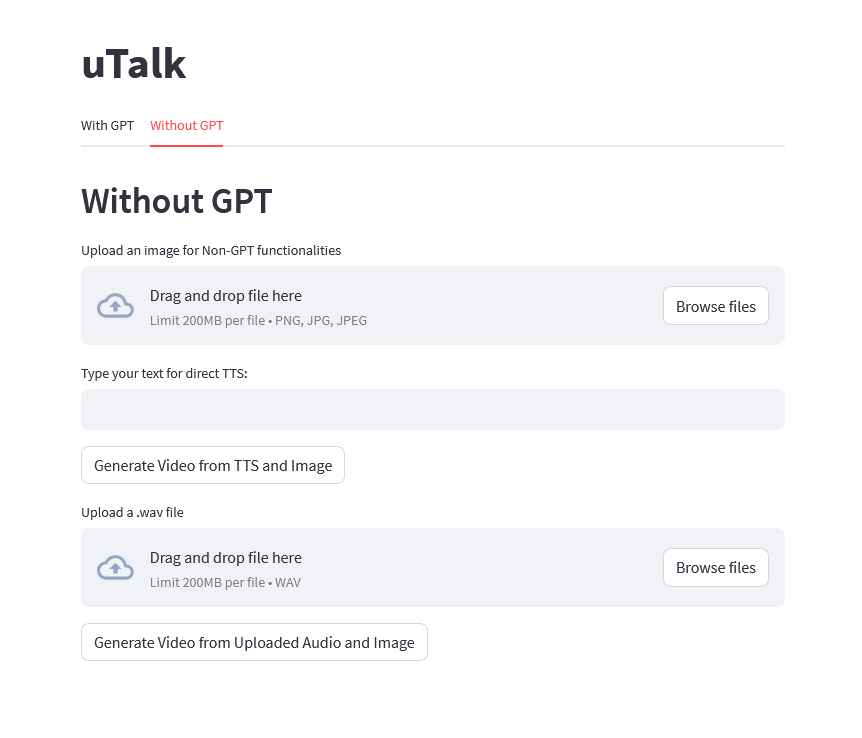}}
    \end{minipage}
    \caption{uTalk consists of two available tabs. The ChatGPT tab \textbf{(left)} generates videos responding to the question provided. On the other hand, the second tab \textbf{(right)} utilizes the input directly to generate video without the use of an LLM for response generation.}
    \label{uTalk_UI}
\end{figure*}

The user interface of our system prioritizes simplicity, ensuring convenience for the user. Fig. \ref{uTalk_UI} demonstrates the User Interface (UI). It contains two tabs depending on the application needed: conversing with an AI assistant or generating videos using the input. Either way, the process starts with an uploaded image, and then the user will be prompted to provide text or audio as input to generate the video. Finally, the generated video is shown on the streamlit web application.

\section{Results and Discussion}

We have performed a series of incremental experiments to enhance the overall performance and integration of the system, ensuring the effectiveness of each method. All experiments use a Windows 11 machine, with an SSD running the operating system (OS), an NVIDIA Geforce RTX 3090Ti, and a 12th Gen i9-12900 Intel CPU.

\subsection{SadTalker Optimization}

First, we removed the tqdm library as it is mainly used for tracking progress, which does not affect the overall result. This removal led to a reduction in the overall time from 40.637 seconds to 39.930 seconds. Second, most code optimization occurred while speeding up facexlib library by importing gfpGAN within our code base, resulting in finding multiple bottlenecks. Firstly, we modified the resizing of the restored face from 512x512 to 256x256 as the overall performance is not affected noticeably, resulting in a much faster inference. Second, with paste-faces-to-input-images, two Gaussian blur functions are removed as they increase the run-time without any noticeable effect on the performance. Finally, as our implementation is forwarded towards real-time performance, we replaced the detection model within the faceRestorerHelper from retinaface-resnet50 to retinaface-mobile0.25. These changes resulted in a drastic improvement in the speed, going from 39.930 seconds to 31.182 seconds. Third, the initial model saves all intermediate values until it generates the video, then deletes it all. However, instead of that, we changed the code not to save any of the values except the final output videos. Even though it should result in an improvement, we saw a minor increase in run-time. Finally, based on our initial bottleneck profiling, mimsave is a time-consuming process; thus, we modified the code to integrate openCV instead of mimwrite to output the final generated video. This change improved the overall performance from 31.438 seconds to 29.385 seconds. So all in all, Table \ref{speedcomparison} summarized the comparison, showing that we could reduce the run-time by 27.69\%.

\subsection{FPS importance}

A subjective study is conducted across a diverse group to quantify the perceived smoothness of AI-generated videos with varying FPS. This study helps in understanding the threshold between smoothness and choppiness. This study involved 29 participants who watched nine videos with FPS ranging from 16 to 24. The order of the videos has been randomized to eliminate potential bias. The participants of this study consisted of 55.2\% of GenZ age demography (1997-2012) and 44.8\% of Millennial age demography (1981-1996). Table \ref{FPS_subjective} shows the results based on the subjective study. We want to note that one participant could not rate the 23FPS video, so that rating was excluded. A variety of conclusions are obtained these results. Generally, videos with less than 18 FPS were less smooth. All the others performed reasonably well, especially the videos with 20, 23, and 24 FPS. Although the study was designed to observe the impact of FPS, it is worth noting that the videos with 17 and 22 FPS had lower mean scores than the others. These low scores might be due to generated head movements or the participants' preferences. The study suggests that 20FPS generated videos are comparable to 23FPS, showing that it is possible to provide a 14.88\% speedup compared to the original 25FPS as shown in Table \ref{fps_speedup} without affecting the perceived quality.

\subsection{Framework Integration}

The integration of Sadtalker with Streamlit yielded notable gains in computational speed, particularly in the context of parallel operations. This parallelism comes from modularizing our framework into initialization and generation modules. We evaluated the performance based on a five-time run by running both the initial framework (without initialization) and with initialization. By pre-loading and initializing the models at the onset of the Streamlit application, we saw a marked reduction in average video generation time. The time taken to produce a 7-second video with 25FPS was reduced from 33.19 seconds with a standard deviation of 1.14 seconds to 29.94 seconds with a standard deviation of 1.11 seconds. This shows a 9.8\% reduction in run-time, enhancing the user experience.

\begin{table}
\caption{Subjective study shows the average rating and its standard deviation (SD) of 9 videos from 1 (very poor) to 5 (very good) based on the smoothness of the AI-generated videos.}
\resizebox{\columnwidth}{!}{
\begin{tabular}{|l|l|l|l||l|l|l|l|}
\hline
\textbf{FPS} & \textbf{Mean Score (SD)} & \textbf{Min} & \textbf{Max} & \textbf{FPS} & \textbf{Mean Score (SD)} & \textbf{Min} & \textbf{Max} \\ \hline

16 & 2.83 (1.10) & 1 & 5 & 
21 & 3.14 (1.03) & 1 & 5 \\
\hline

17 & 2.21 (1.21) & 1 & 5 & 
22 & 2.97 (1.15) & 1 & 5 \\
\hline

18 & 3.48 (0.87) & 1 & 5 & 
23 & 3.71 (1.15) & 1 & 5 \\
\hline

19 & 2.83 (1.10) & 1 & 5 & 
24 & 3.62 (1.08) & 1 & 5 \\
\hline

20 & 3.66 (1.01) & 1 & 5 &
-  & - & - & - \\
\hline

\end{tabular}}
\label{FPS_subjective}
\end{table}

\section{Conclusion}

This work proposes uTalk, a framework that integrates an optimized version of SadTalker with various algorithms, such as Whisper API, ChatGPT, and TTS via Azure Cognitive Services, all within a UI hosted on Streamlit. The interface comprises two primary functionalities: First, generating a response video through ChatGPT as the user provides an audio or text input. Second, instead of utilizing ChatGPT, we generate a video of the avatar speaking the input. Throughout this work, we explore how SadTalker is optimized to enhance the overall run-time, benefitting the user experience. Additionally, we conducted a subjective study to find the optimal FPS of AI-generated videos, concluding that using 20FPS will reduce the overall run-time without a noticeable effect on video quality. Ultimately, the user interface combines all these features to guarantee a seamless experience for conversing with a digital twin or creating content.

\begin{table}
\caption{Run-time comparison based on the frames per second (FPS)}
\resizebox{\columnwidth}{!}{
\begin{tabular}{|l|l|l||l|l|l|}
\hline
\textbf{Model} & \textbf{FPS} & \textbf{Run-time (seconds)} & \textbf{Model} & \textbf{FPS} & \textbf{Run-time (seconds)} \\ \hline
Experiment 1 & 25 & 29.385 $\pm$ 0.284 & Experiment 6 & 20 & 25.041 $\pm$ 0.104 \\ \hline
Experiment 2 & 24 & 28.525 $\pm$ 0.199 & Experiment 7 & 19 & 24.196 $\pm$ 0.217 \\ \hline
Experiment 3 & 23 & 27.833 $\pm$ 0.137 & Experiment 8 & 18 & 23.026 $\pm$ 0.131 \\ \hline
Experiment 4 & 22 & 26.842 $\pm$ 0.189 & Experiment 9 & 17 & 22.134 $\pm$ 0.131 \\ \hline
Experiment 5 & 21 & 25.899 $\pm$ 0.301 & Experiment 10 & 16 & 21.241 $\pm$ 0.137\\ \hline
\end{tabular}}
\label{fps_speedup}
\end{table}

\bibliography{references}
\bibliographystyle{IEEEtran}

\vspace{12pt}

\end{document}